# SYNTHESIZING DYSARTHRIC SPEECH USING MULTI-SPEAKER TTS FOR DYSARTHRIC SPEECH RECOGNITION

*Mohammad Soleymanpour[1], Michael T. Johnson[1], Rahim Soleymanpour[2], Jeffrey Berry[3]*

[1] Electrical and Computer Engineering, University of Kentucky, Lexington, KY USA 40506
[2] Department of Biomedical Engineering, University of Connecticut, Storrs, CT USA 06269,
[3] Speech Pathology and Audiology, Marquette University, Milwaukee, WI USA 53201

{m.soleymanpour, mike.johnson}@uky.edu , rahim.soleymanpour@uconn.edu , jeffrey.berry@marquette.edu

## ABSTRACT

Dysarthria is a motor speech disorder often characterized by reduced speech intelligibility through slow, uncoordinated control of speech production muscles. Automatic Speech recognition (ASR) systems may help dysarthric talkers communicate more effectively. To have robust dysarthria-specific ASR, sufficient training speech is required, which is not readily available. Recent advances in Text-To-Speech (TTS) synthesis multi-speaker end-to-end TTS systems suggest the possibility of using synthesis for data augmentation. In this paper, we aim to improve multi-speaker end-to-end TTS systems to synthesize dysarthric speech for improved training of a dysarthria-specific DNN-HMM ASR. In the synthesized speech, we add dysarthria severity level and pause insertion mechanisms to other control parameters such as pitch, energy, and duration. Results show that a DNN-HMM model trained on additional synthetic dysarthric speech achieves WER improvement of 12.2% compared to the baseline, the addition of the severity level and pause insertion controls decrease WER by 6.5%, showing the effectiveness of adding these parameters. Audio samples are available at
https://mohammadelc.github.io/SpeechGroupUKY/

## 1. INTRODUCTION

Dysarthria is a motor speech disorder, often caused by traumatic injury or neurological disease [1], which decreases speech intelligibility through slow, uncoordinated control of speech production muscles. People with moderate and severe levels of dysarthria are less able to communicate with others through speech due to poor intelligibility [2].

Talkers with dysarthria may exhibit imprecise articulation, irregularities of vocal pitch and quality, atypical nasal resonance, slow and inconsistent speaking rate, inconsistent pauses, as well as altered linguistic stress and speech sound timing [3]. Although individuals with dysarthria may have the cognitive and language abilities to formulate communication, they may not be able to reliably plan and execute the muscle control needed for sufficiently intelligible speech. Speech technologies such as ASR have the potential to improve the effectiveness of dysarthric speakers' communication if accurate dysarthric-specific ASR models are available.

There are a few publicly available dysarthric speech datasets, including TORGO [4], UASpeech [2] and Nemours [5]. However, none of these datasets are designed for speech recognition, and using them to support ASR is challenging. Because there is not an adequate amount of conversational speech in these datasets, ASR systems trained with these are often less robust. Furthermore, modern ASR methods assume that training data include a sufficiently large set of speakers to adequately capture enough inter-speaker variability, but these dysarthric datasets all have a relatively small number of speakers and are not sufficient for an ASR system to capture speaker variability.

Recent progress in end-to-end TTS systems such as Tacotron [6, 7], FastSpeech [8, 9], Deep-Voice [10] support synthesized speech with high quality and naturalness with varying prosody. These improvements in synthesizing speech inspired us to attempt synthesis of realistic dysarthric speech for ASR training data augmentation. Such neural speech synthesizers have been used to generate new utterances for ASR application for low resource languages [11-15]. Multi-speaker speech synthesis systems can learn prosody characteristics, speaker and style variation extracted from the training set, and can use speaker embeddings to generate speech in a variety of speaker styles [13-15]. This allows for generation of relatively large amounts of the high-quality synthesized speech across a range of speaker characteristics and speaking styles.

In this paper, we propose a method based on Multi-talker neural TTS to synthesize dysarthric speech to enhance the results of dysarthric ASR. In addition to traditional prosodic variables such as speech rate, energy, and pitch, we add two new variables to control dysarthric severity and extent of pause insertion. These parameters enable us to generate a broad range of synthesized speech to improve the training of dysarthric ASR systems. To assess the effectiveness of the synthetic speech, we evaluate the Deep Neural Network-Hidden Markov Model (DNN-HMM) models with and without augmented speech. Experiments are carried out using the proposed approach on the TORGO dataset.

## 2. METHODOLOGY

For the baseline synthesis model, we modified FastSpeech2 [9] and a recent variant [8] to synthesize dysarthric speech. Figure 1 shows the main block diagram of the proposed method. In the modified version of the FastSpeech2 the energy, pitch and forced-alignment duration [16] of each speaker's utterances are incorporated into the phoneme hidden sequence through a "variance adaptor" module, resulting in more controllability of these prosodic parameters.

The multi-talker variant of FastSpeech2 decoder works like a voice conversion system, making it a multi-talkers TTS [8] capable of generating speech in a wide range of speaking styles. This is a useful capability for speech synthesis for data augmentation because it allows generation of a robust set of training data.

The prosodic characteristics of dysarthric speech greatly differs from typical speech, specifically at moderate and high severity levels. One significant difference between dysarthric and typical speech is that the speaking rate is often substantially slower for talkers with dysarthria [2, 3]. However, this reduced speaking rate is often not consistent throughout the utterance. In addition, dysarthric vocal excitation may be unstable because many individuals with dysarthria cannot effectively control vocal fold closure and vibration. This may cause inconsistent vocal quality and pitch throughout an utterance.

### 2.1 Synthetic Dysarthric Speech

These differences in speech style and speaking rate significantly depend on the dysarthria severity level of the talkers [17]. To be able to synthesize accurate dysarthric speech, we add a dysarthria severity predictor in the variance adaptor to simulate the characteristics of different severity levels of dysarthric speech. The severity embedding is added as an input to the variance adaptor before the pitch/energy/duration predictors. This allows the system to detect the relative characteristics of different severity groups, especially duration, pause and voice harshness, and variance of pitch and energy. It also allows additional control of the duration of the speech

Like the duration, pitch, and energy predictors, the dysarthria severity level predictor has a similar model structure which consists of a 2-layer 1D-convolutional network with ReLU activation, each followed by the layer normalization and a dropout layer, and an extra linear layer to project the hidden states into the output sequence [9].

Based on the structure of the TORGO dataset and the amount of data available, speakers are categorized into three dysarthria severity levels: normal, very low/low, and medium and into two intelligibility groups: intelligible and non-intelligible [17]. During training, the TORGO non-dysarthric speakers are used for the Normal category, coefficient 0. Because there is only one speaker in the Low category, Very Low and Low are combined together to form the middle category, coefficient 1. The highest severity level in the dataset, labeled Medium, is used for the third category, coefficient 2.

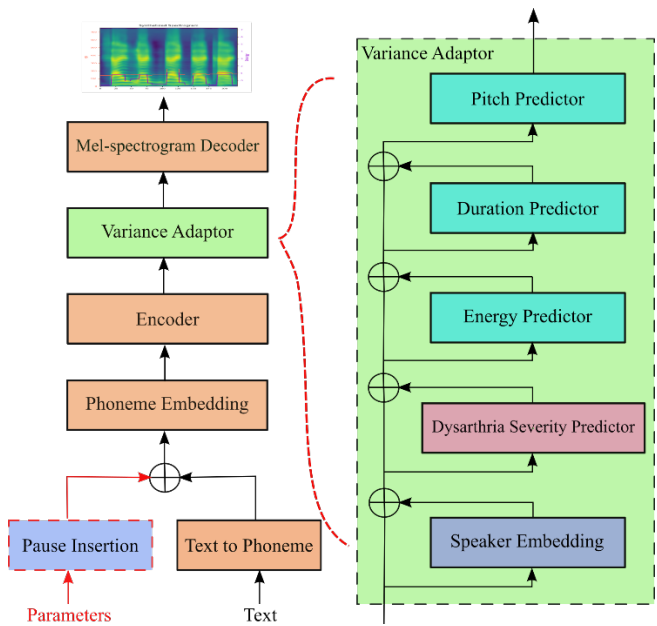

Figure 1. An overview of the proposed architecture

### 2.2 Pause Insertion

Pause is another important indicator of dysarthric speech. Analysis of the TORGO data set indicates that the number of between-word pauses per sentence among typical, very low, low and moderate groups is about 0.26, 0.57, 1.21 and 2.51, respectively. As a ratio to normal speakers, this means that the number of pauses is 120%, 365%, and 865% more frequent among the very low, low and moderate groups in comparison with typical speaker in TORGO dataset. The effect of sentence length on pause duration has also been previously investigated in persons with dysarthria due to Amyotrophic Lateral Sclerosis (ALS) [4]. Their results showed that the pause duration over sentence length for the group with higher severity level was increased by a higher rate in comparison with the group with lower severity level.

Although FastSpeech2 can already synthesize normal pause patterns for a given text, it is not sufficient to represent the patterns in dysarthric speech. To address this issue, we add a binary parameter to control insertion of additional pauses. Although pauses in dysarthric speech sometimes occur between phonemes within a word as well, the current version supports insertion of pauses only between words. To implement this, possible inter-word positions are identified, and then the maximum number of pauses is determined based on the severity level and length of the given sentence. For longer texts or for speakers with a higher dysarthria severity level, the model inserts more pauses. Since many of the sentences in the TORGO dataset are relatively short, there is not enough data to learn a complex model for pause insertion, so a simple model is used. The model uses the number of words in the sentence and the dysarthric severity level to determine the number of pauses to be inserted. Once this is set, the locations of the pauses are chosen randomly at inter-word locations in the sentence. The pause insertion model is shown in the bottom left of the architecture in Figure 1.

For ASR, Pytorch-kaldi [18] is used to train DNN ASR models. A light Gated Recurrent Unit (liGRU) architecture is implemented, trained on fMLLR transformed features. Baseline configuration files provided in the Pytorch-kaldi repository for common speech databases like TIMIT, Librispeech were used as reference and the final architecture was based on experimental results using a small number of training set speakers [20].

## 2. EXPERIMENTAL SETUP

FastSpeech2 contains 4 feed-forward transformer blocks in the encoder and mel-spectrogram decoder. The decoder generates an 80-dimensional mel-spectrogram from hidden state. The size of phoneme embedding is 256 in our implementation. We train the adjusted model with a GeForce RTX 2080 Ti on the TORGO [4] dataset, containing 8 dysarthric speakers and 7 normal speakers. This dataset consists of non-words (which are excluded in this experiment), short words, restricted and non-restricted sentences. Two types of microphones were used in this dataset, a head-mounted microphone as well as an array of 8 microphones. Dysarthric speakers are categorized into three dysarthria severity levels, very low, low, and medium and into two groups for intelligibility, intelligible and non-intelligible [17]. The number of utterances for each dysarthric talker averages 700; whereas for normal speakers the average is 1560 [2].

After training the TTS models, the text in TORGO is used to synthesize additional dysarthric speech. The effect of the synthesized speech is evaluated by implementing two experiments on speech recognition application.

In the first experiment, we focus on the effect of the severity predictor and pause insertion which is the main contribution of this paper. Synthesized speech for augmentation is synthesized with three different severity coefficients of 0.0, 1.0, and 2, with the pause insertion turned on. Pitch, energy and duration coefficients are fixed at 1.0. The number of augmented sentences was three times that of the original TORGO dataset.

For the second experiment, we synthesize a wider range of dysarthric speech for augmentation across all controllable parameters. Parameters for pitch, energy, duration, as well as severity level were varied across a range with pause insertion activated as shown in Table 1 below. The number of augmented sentences was ten times that of the original TORGO dataset.

Table 1- The prosody coefficients for synthesizing dysarthric speech in the two experiments

| Coef. | Baseline | Exp. 1 | Exp. 2 |
|---|---|---|---|
| Pitch | - | 1.0 | [0.1, 0.6, 1.2, 1.75] |
| Energy | - | 1.0 | [0.1, 1.0, 2.0] |
| Duration | - | 1.0 | [1.0, 1.3, 1.6, 1.8] |
| Severity level | - | [0.0, 1.0, 2.0] | [0.0, 1.0, 2.0] |
| Pause insertion | - | Yes | Yes |
| Total utterance | ~16000 | ~×3 | ~×10 |

The synthesized speech is applied for training the DNN-HMM model with light bidirectional GRU [18] architecture, with five layers containing 1024 cells each, activated by Relu activation function and dropout of 0.2. The number of epochs was 10 to 12 to achieve the best result of each experiment. The architecture applies monophone regularization [21]. A multi-task learning procedure was applied using two SoftMax classifiers, one estimating context-dependent states and the second one predicting monophone targets [20].

For testing, a leave-one-speaker-out cross-validation procedure was applied across the original TORGO dataset.

## 4. RESULTS AND DISCUSSION

Before evaluating the performance of the synthetic data augmentation on dysarthria-specific DNN-HMM speech recognition, we first review and assess the quality of the synthesized dysarthric speech itself. Figure 2 shows the synthesized speech of speaker MC04 for the input text "We would like to play volleyball" for severity level of 0, 1 and 2, respectively. Pitch, energy and duration coefficients are the same across the various severity levels shown here. As indicated in Figure 1, synthesized speech duration increases with increasing severity level. Duration is one of the key indicators of different levels of severity. Other parameters, including harshness, blurred and unintelligibility are also synthesized and can be evaluated on the provided demo web page[1].

To evaluate the results of the two data augmentation ASR experiments, Word Error Rate (WER) was calculated for each test speaker. Table 2 shows the WER of the two experiments along with the baseline and compares them with the results of the best models of two other published works preformed on the same TORGO dataset using hybrid speech recognition models [22, 23].

---
[1] https://mohammadelc.github.io/SpeechGroupUKY/

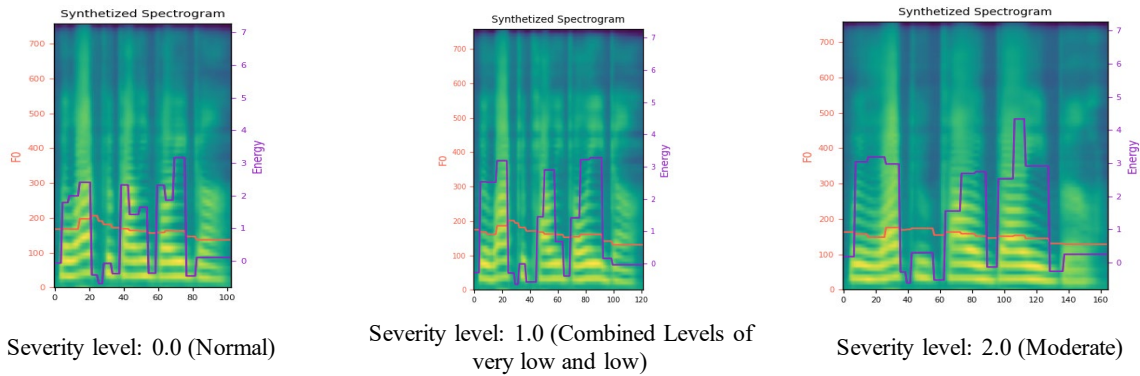

Figure 2- Effect of dysarthria severity coefficients in synthesizing dysarthric speech for speaker MC04

Results show that the WER performance of the baseline is similar to that of the two comparison methods for the lowest few severity levels, and slightly better for the highest ("medium") severity. The average WER across all speakers is 44.5%, 56.2% and 43.3% for our baseline, [22] and [23], respectively.

In the first experiment that only used severity synthesis and pause insertion, the synthesized speech used for augmenting ASR training improved the performance of DNN-HMM model for each speaker except M03, which declined slightly. Average WER performance across all speakers improved from 44.5% to 41.6%. The second experiment with additional prosody variance and data augmentation showed further performance improvement, with individual improvement for all 8 speakers in the dataset.

Average WER performance across all speakers improved from 44.5% to 39.2%. On average, the first and second experiments reduced WER by 6.5 %, 12.2% with the respect to the baseline, respectively.

Table 2- WER of each test speaker for the two augmentation experiments: Exp.1 included augmented speech across 3 severities with pause insertion, and Exp. 2 included augmented speech across severity, pause, pitch, energy, and duration.

| Severity Level | Test Spk | WER (%) | | | | |
|---|---|---|---|---|---|---|
| | | Baseline | Exp. 1 | Exp. 2 | [22] | [23] |
| Very low | F04 | 16.8 | 16.3 | 14.5 | 18.3 | 13.1 |
| | M03 | 10.9 | 12.7 | 10.7 | 18.2 | 17.7 |
| Low | F03 | 46.6 | 39.3 | 36.8 | 44.2 | 39.1 |
| Moderate | F01 | 58.3 | 52.4 | 50.4 | 71.5 | 39.6 |
| | M01 | 55.4 | 51.3 | 50.3 | 69.3 | 62.2 |
| | M02 | 44 | 43.1 | 38.4 | 70.9 | 42.9 |
| | M04 | 65.8 | 64.2 | 62 | 79.9 | 69.0 |
| | M05 | 58.2 | 53.6 | 49.6 | 77.2 | 62.6 |
| Overall Average | | 44.5 | 41.6 | 39.2 | 56.2 | 43.3 |

To summarize the effect of the proposed approaches as a function of the level of severity of the dysarthric speech, Table 3 shows the average WER for speakers at the different dysarthria severity levels. This shows that augmentation using synthetic speech at three dysarthria levels with pause insertion improved the WER of each severity level on average except for the group with the low severity. Augmentation using synthetic speech at three severity levels plus pause insertion, further varying energy, pitch, and duration improved WER across all severity levels.

Table 3- WER of each severity level for the two augmentation experiments.

| Severity level | baseline | Exp. 1 | Exp. 2 | Improvement | |
|---|---|---|---|---|---|
| | | | | Exp.1 | Exp.2 |
| Very Low | 13.8 | 14.5 | 12.6 | -4.7% | 9% |
| Low | 46.6 | 39.3 | 36.8 | 7.3% | 21% |
| Moderate | 56.3 | 52.9 | 50.1 | 6% | 11% |
| All | 44.5 | 41.6 | 39.2 | 6.5% | 12.2% |

## 5. CONCLUSION

In this paper, we have modified a neural multi-talker TTS by adding a dysarthria severity level coefficient and a pause insertion model to synthesize dysarthric speech for varying severity levels. We evaluate its effectiveness for data augmentation of training data for dysarthria-specific speech recognition. Results are shown for two different experiments: the first includes augmented speech across 3 severities with pause insertion, and the second includes augmented speech across severity, pause, pitch, energy, and duration. Overall results on the TORGO database demonstrate that using dysarthric synthetic speech to increase the amount of dysarthric-patterned speech for training has significant impact on the dysarthric ASR systems. A demonstration web page with audio results of the synthesis is available at https://mohammadelc.github.io/SpeechGroupUKY/ .

## 6. ACKNOWLEDGEMENT

This work was supported by National Institutes of Health under NIDCD R15 DC017296-01.